# RDBNorma: - A semi-automated tool for relational database schema normalization up to third normal form.

Y. V. Dongare[1], P. S. Dhabe[2] and S. V. Deshmukh[3]

[1]Department of Computer Engg., Vishwakarma Institute of Info.Technology, Pune, India
yashwant_dongre@yahoo.com

[2]Department of Computer Engg., Vishwakarma Institute of Technology, Pune, India
dhabeps@gmail.com

[3]Department of Computer Engg., Pune Vidhyarthi Griha's College of E&T, Pune, India
supriya_vd2005@yahoo.com

***Abstract—*** *In this paper a tool called RDBNorma is proposed, that uses a novel approach to represent a relational database schema and its functional dependencies in computer memory using only one linked list and used for semi-automating the process of relational database schema normalization up to third normal form. This paper addresses all the issues of representing a relational schema along with its functional dependencies using one linked list along with the algorithms to convert a relation into second and third normal form by using above representation. We have compared performance of RDBNorma with existing tool called Micro using standard relational schemas collected from various resources. It is observed that proposed tool is at least 2.89 times faster than the Micro and requires around half of the space than Micro to represent a relation. Comparison is done by entering all the attributes and functional dependencies holds on a relation in the same order and implementing both the tools in same language and on same machine.*

***Index Terms—*** *relational databases, normalization, automation of normalization, normal forms.*

## 1. INTRODUCTION

Profit of any commercial organization is depends on its productivity and quality of the product. To improve the profit they need to increase productivity without scarifying quality. To achieve this, it is necessary for organizations to automate the tasks involved in the design and development of their products.

From past few decades relational databases proposed by Dr. Codd [1] are widely used in almost all commercial applications to store, manipulate and use the bulk of data related with a specific enterprise, for decision making. Detail discussion on relational database can be found in [2]. Their proven capability to manage the enterprise in a simple, efficient and reliable manner increased a great scope for software industries involved in the development of relational database system for their clients.

Success of relational database modeled for a given enterprise is depending on the design of relational schema. An important step in the design of relational database is "Normalization", which takes roughly defined bigger relation as input along with attributes and functional dependencies and produces more than one smaller relational schema in such a way that they will be free from redundancy, insertion and deletion anomalies [1]. Normalization is carried out in steps. Each step has a name First normal form, second normal form and third normal form represented shortly with 1NF, 2NF and 3NF respectively. First three normal forms are given in [1] [2]. Some other references also help to understand the process of normalization [3], [4], [5], [6], [7], [8] and [9].

We found some papers very helpful about normalization. This paper [10], explains 3NF in an easiest manner. The 3NF is defined in different in equivalent ways in various text books again their approach is non-algorithmic. They have compared definitions of 3NF given in various text books and present it an easy way so that students can understand it easily. They have also claimed that an excellent algorithmic method of explaining 3NF is available which is easy to learn and can be programmed. Ling et.al [11], proposed an improved 3NF, since Codds 3NF relations may contain Superfluous (redundant /

unnecessary) attributes resulting out of transitive dependencies and inadequate prime attributes. In their improved 3NF guarantees removal of superfluous attributes. They have proposed a deletion normalization process which is better than decomposition method. Problems related with functional dependencies and algorithmic design of relational schema are discussed in [12]. They have proposed a tree model of derivation of functional dependency from other functional dependencies, a linear time algorithm to test if a functional dependency is in closure set and quadratic time Bernstein's third normal form. Concept of multivalued dependency [13] which is generalization of functional dependency and 4NF which is used to deal with it is defined in [3]. This normal form is stricter as compared to Codd's 3NF and BCNF. Every relation can be decomposed into family of relations into 4NF without loss of information. The 5NF also called as PJ/NF is defined in [14]. This is an ultimate normal form where only projections and joins operations are considered hence called PJ/NF. It is stronger than 4NF. They have also discussed relationship between normal forms and relational operators. In [15] a new normal form is defined called DK/NF. That focuses on domain and key constraints. If a relation is in DK/NF then it has no insertion and deletion anomalies. This paper defines concept of domain dependency and key dependency. A 1NF relation is in DK/NF if every constraint is inferred from domain dependencies and key dependencies. This paper [16] proposed a new normal form between 3NF and BCNF. It has qualities of both. Since 3NF has inadequate basis for relational schema design and BCNF is incompatible with the principle of representation and prone to computational complexity. [17] proposed new and fast algorithms of databse normalization.

## 2. RELATED WORK

Normalization is mostly carried out manually in the software industries, which demand skilled persons with expertise in normalization. To model today's enterprise we require large number of relations, each containing large number of attributes and functional dependencies. So, generally, more than one persons need to be involved in manual process of normalization. Following are the obvious drawbacks of normalization carried out manually.
1. It is time consuming and thus less productive:- To model an enterprise a large number of relation containing large number of attributes and functional dependencies may be required.
2. It is prone to errors: - due to reasons stated in 1.
3. It is costly: - Since it need skilled persons having expertise in Relational database design.

To eliminate these drawbacks several researchers already tried for automation of normalization by proposing new tools/methods. We have also seen a US patent [18], where a database normalizing system is proposed. This system takes input as a collection of records already stored in a table and by observing a record source it normalizes the given database. Hongbo Du and Laurent Wery [19] proposed a tool called *Micro,* which uses two linked lists to represent a relation along with its functional dependencies. One list stores all the attributes and other stores functional dependencies holds on it. Ali Ya zici, et.al [20] proposed a tool called *JMathNorm*, which is designed using inbuilt functions provided by Mathematica and thus depend on Mathematica. This tool provides facility to normalize a given relation up to Boyce-codd normal form including 3NF. Its GUI interface is written in Java and linked with Mathematica using Jlink library. Bahmani et. al [21], proposed an automatic database normalization system that creates dependency matrix and dependency graph. Then algorithms of normalization are defined on them. Their method also generates relational tables and primary keys.

In this work, we also found some good tools specifically designed for learning/teaching/understanding the process of normalization, since the process is difficult to understand, dry and theoretical and thus it is difficult to motivate the students as well as researchers. Maier [22], also claimed that the theory of relational data modeling (normalization) tend to be complex for average designers. CODASYS, a tool that helps new database designer to normalize with consultation [23]. A web based, client-server, interactive tool proposed in [24], called LBDN (Learn DataBase Normalization) that can provide hands-on training to students and some lectures for solving assignments. It represents attributes, functional dependencies and keys of a relation in the form of sets, stored as array of strings. A similar tool is proposed in [25], which is also web based and can be used for system analysis and design and data management courses. Authors of this tool claimed that this tool is having a positive impact on students.

Our tool *RDBNORMA* uses only one linked list to represent a relation along with functional dependencies holds on it and thus a novel approach that requires less space and time as compared to *Micro*. Our proposed system *RDBNORMA* works at schema level

This paper is a sincere attempt to develop a new way of representation of a relational schema and its functional dependencies using one linked list thus saving memory and time both. This representation helps to automate the process of relational database schema normalization using a tool which works at schema level, in a faster manner. This work reduces the drawbacks of manual process of normalization by improving productivity.

Remaining parts of the paper are organized as follows. Section 3 describes signally linked list node structure used to represent a relation in computer memory along with Functional Dependencies ( FD's). Algorithms for storing a relations and their FD's are described in section 4. Section53 demonstrates a real world example for better understanding of algorithms to store a relation. Design constraints are discussed section 6. Section 7 elaborates algorithm for 1NF. Algorithm of minimal cover is discussed in Section 8. Algorithm of 2NF and 3NF are discussed in Section 9 and 10, respectively. Standard relational schemas used for experimentation are discussed in Section 11. Experimental results and comparison is done in Section 12. Conclusions based on empirical evidences are drawn in section 13 and references are cited at the end.

## 3. NODE STRUCTURE USED FOR REPRESENTATION OF A RELATION IN RDBNORMA

### A. Problems in representing a relation

At the initial stage we have decided to represent a relation using a signally linked linear list. But we need to address two things for it; first, how to store attributes? and the second, how to store FD's?. We have decided to store one attribute per linked list node as in Micro [Du and Wery, 1999]. But using a separate linked list for storing all the FD's holds on that relation as in Micro [Du and Wery, 1999], according to us, although it is convenient but not optimal. Thus we have decided to incorporate the information about the FD's in the same linked list and come up with following design of the node structure. Again in what order we have to inter attributes into a linked list? Need to be finalized. We have decided to enter all the prime attributes first and then non prime ones. This specific order helps us to get determiners of non prime attributes since they will be already entered in linked list.

### B. Node structure

The node structure used to represent a relation need to have ten fields as shown in Fig. 1.

| *attribute_name* |
| --- |
| *attribute_type* |
| *determiner* |
| *nodeid* |
| *determinerofthisnode1* |
| *determinerofthisnode1* |
| *determinerofthisnode1* |
| *determinerofthisnode1* |
| *keyattribute* |
| *ptrtonext* |

Fig. 1. Linked list Node structure.

The description and use of these fields are as follows.

International Journal of Database Management Systems ( IJDMS ), Vol.3, No.1, February 2011

1. *attribute_name*:- This field is used to hold the attribute name. It allows underscores and special character and size can at least 50 characters or more based on the problem at hand . We assume unique attribute names within a given databases, but two relations can have same attribute names for referential integrity constraints like foreign keys.
2. *attribute_type:*- This field is used to hold type of the attribute and will hold *-for multivaled attribute, 1 for atomic attribute. It will be of size 1 character long.
3. *determiner*: - Determiner is a field which takes part in left hand side of FD. This field indicates whether this attribute is determiner or not and of binary valued a size of 1 character will be more than sufficient. If this filed is set to 1 indicates that this attribute is a determiner otherwise it is dependant.
4. *nodeid*:- It is a node identifier ( a unique number ) assigned to each newly generated node and is stored inside the node itself . This number can be generated by using a *NodeIDCounter*, which needs to be reset for normalizing a new database. When new node is added on a linked list *NodeIDCounter* will be incremented by 1. A sufficient range need to be defined for this *nodeid* e.g. [0000-9000]. Upper bound 9000 indicate that a database can have at most 9000 attributes. Size of this filed is based on the range defined for this attribute.
5-8. *determinerofthisnode1, determinerofthisnode2, determinerofthisnode3 and determinerofthisnode4:-* These fields hold all the determiners of this attribute assuming that there can be at the most 4 determiners of an attribute, for example as shown in following FD's an attribute E has 4 determiners ABCD, GH, AH and DH.

$$A, B, C, D \rightarrow E$$
$$G, H \rightarrow E$$
$$A, H \rightarrow E$$
$$D, H \rightarrow E$$

A Determiner can be composite or atomic. E.g. Consider this node represents an attribute C and we have AB->C and D->C then the two determiners of C are (A,B) and (D) and thus their *nodeid*'s will be stored in *determinerofthisnode1* and *determinerofthisnode2* and *determinerofthisnode3* and *determinerofthisnode4* will be hold NULL. Each of this field can hold at most 4 *nodeid*'s, it means that left hand side of a FD's can not have more than 4 attributes. To illustrate use of these fields consider following set of FD's for a dependant attribute H.

$$A, B, C, D \rightarrow H$$
$$E, F \rightarrow H$$
$$G \rightarrow H$$

If *nodeid's* of attribute A, B, C, D, E, F and G are 100, 101, 102, 103, 104, 105, and 106 respectively then determiners fields of node representing attribute H is as shown in Fig. 2, if these FD are entered in the same order as shown.

| ... | 103 | NULL | NULL | NULL | ... |
|---|---|---|---|---|---|
| | 102 | NULL | NULL | NULL | |
| | 101 | 105 | NULL | NULL | |
| | 100 | 104 | 106 | NULL | |

Fig.2. Determiner fields of attribute H.

9. *keyattribute*:- This is a binary filed and hold 1 if this attribute is taking participation in primary key else it is 0. Size of 1 character is sufficient for this purpose.
10. *ptrtonext*:- This filed hold pointer (link) to next node and will be NULL if this is the last node on the list.

## 4. ALGORITHMS FOR STORING A RELATION AND ITS FUNCTIONAL DEPENDENCIES (FD'S)

This tool needs three algorithms for doing its work. Representing a relation using linked list in computer memory involve adding a new node for each attribute and for adding each separate FD's we need to update information in nodes representing those attributes participating in this FD's. For adding all the attributes of a relation we need algorithm *AddNewAttribute*, which uses another algorithm *CreateNewNode* internally. User has to find out composite attributes and need to be replaced by their atomic attribute components, thus 1NF can be achieved at the attribute entry level.

*A. Algorithm for adding a new attribute on linked list.*

Algorithm *AddNewAttribute ( listptr, x, NodeIDCounter)*

This algorithm adds a new attribute node with attribute name x on linked list using *a nodeid= NodeIDCounter* value. Name of the relation is used as *listptr*, which points to the first node on that linked list. If *listptr=NULL* means list is empty we need to create first node for that relation. It uses function *CreateANewNode( )*, which creates a new node and returns its link. This algorithm uses two variable pointers p and q. This algorithm is described in Fig. 3.

*B. Algorithm for creating a new node.*
Algorithm *CreateANewNode( )*
This algorithm returns a list node pointer. Operator new will create a new node of struct node type as shown in Fig. 1 and will return its pointer. It is as shown in Fig. 4.

```
Input: pointer to list listptr (relation name if it at least one attribute node is created),
       x a new attribute to be added on list, counter value to set nodeid of this new node.
Output: Returns nothing, but adds new attribute node on linked list.

            BEGIN
              / * listptr   is relation   name   * /
              If  listptr   = =  NULL   then
              / * means   if list is empty then   create   a new   node
                                 and set its pointer to   listptr.   * /
                    p  =  CreateANew   Node( );
                    listptr   =  p;
              else
              / * if listptr   is not null   * /
                    q  =  listptr;
                    w hile  (q → ptrtonext   ! = NULL)
                          q  =  q →  ptrtonext;
              / * Now   q will point to   the last node   on the list.   * /
                    p  =  CreateANew   Node( );
              endif
                    p → attributen   ame  =  x;
                    p → nodeid   = NodeIDCoun   ter;
                    print("Is   x a determiner   ?")
                    If  YES
                         p → determiner   = 1;
                    else
                         p → determiner   = 0;
                    end
                    print("Is   x a key   attribute?   ")
                    If  YES
                         p → keyattribu   te = 1;
                    else
                         p → keyattribu   te = 0;
                    end
                    print ("What   kind of attribute   x is?   Multivaled   - *, Atomic   - 1");
                    p → attributet   ype  =  either   * /1;
                    p → ptertonext   = NULL   :
                    q → ptrtonext   = p;
            END
```

Fig. 3. Algorithm for adding a new node on linked list.

```
Input: - None
Output: - It returns a pointer to newly created
          node.
BEGIN
    q = new (struct node type)
    q → determinerofthisnode1 = NULL;
    q → determinerofthisnode2 = NULL;
    q → determinerofthisnode3 = NULL;
    q → determinerofthisnode4 = NULL;
    return (q)
END
```

Fig.4. Algorithm to create a new node.

*C. Algorithm for adding a new functional dependency of a relation in its linked list.*

Algorithm *AddAFD (determiner,dependant, listptr)*

This algorithm assumes that The functional dependency set it is taking into account is a minimal cover, which is having minium number of FD's and no redundant attribute. Since 2NF and 3NF algorithms work heavily on FDs using minimal cover make them more efficient. Thus each FD's has exactly one attribute towards its right hand side. This algorithm takes input as one FD at a time containing composite or atomic determiner (left hand side of FD)of a single dependent attribute and set this information in the node structure of that dependent by taking into account the *nodeids* of its determiner nodes. E.g. Consider a FD, $AB \rightarrow C$ then determiner1 string of node representing attribute C will hold nodeids of A and B and determiner2, determiner3 and determiner4 will be set to NULL. An attribute can have at most 4 determiners may be composite or atomic since only 4 fields named *determinerofthisnode1, determinerofthisnode2, determinerofthisnode3 ,and determinerofthisnode4 are used.* It is shown Fig. 5.

There will be no problem in finding *nodeid's* of determiners, since we have imposed an order in which attributes need to be entered is that all the prime attributes need to be entered first, then all the attributes which are nonprime and determiners of some attributes and at last all those attributes which are non-prime and non determiners.

## 5. AN EXAMPLE OF STORING A REAL WORLD RELATION AND ITS FUNCTIONAL DEPENDENCIES USING ONE LINKED LIST

This section describes an example of representing a real word relation and its FD's using a signally linked list for better understanding of algorithms discussed above. Consider a relation employee taken from [9] containing *e_id* as primary key *e_s_name* as employee surname, *j_class* indicating job category and CHPH representing charge per hour. This relation and all FD's holds on it are shown below.

$$\text{Employee} \equiv (e\_id, \ e\_s\_name, \ j\_class, \ CHPH)$$
$$e\_id \rightarrow e\_s\_name, \ j\_class, \ CHPH \quad (1)$$
$$j\_class \rightarrow CHPH \quad (2)$$

Initially a new and first node will be created for the prime attribute *e_id*. Let that *NodeIDCounter* is set to 001. Then a node for *e_id* attribute will be created and is as shown in Fig. 6 and will be pointed by a pointer Employee (name of the relation).

The second field in Fig.6 is set to 1, since e_id is an atomic attribute. Third field is set to 1, since e_id is a determiner. Fourth field is set to 001, since it is the *nodeid* of this node. Remaining four fields are set to *NULL*, indicating that each cell of this field is set to *NULL*. The ninth field is set to 1, since *e_id* is a key attribute. The last attribute is set to *NULL* indicating it is the last node on the list.

*Input:* Names of *determiner1* to *determiner4* and a dependent attribute name extracted
       from a FD.
*Output:* Updated linked list with the new FD information added on it.

    BEGIN
       Repeat  for each  FD  holds  on this  relation.
       Step1.  Find  the node  pointer  of dependant  node  using  linked  list pointer  listptr.
             Let this   pointer   be  p.
       Step 2. Find  first, all NULL  determiner  ofthisnode   out of determiner  ofthisnode  1,
             determiner  ofthisnode  2, determiner  ofthisnode  3 and determiner  ofthisnode  4.
             If such a field is not found it means that all the four determiner s of this dependent
             are already  been  filled  and there  is no room  to accommodat  e fifth  determiner   so
             report  failure  and  halt.  Since  this  tool  assume  a maximum   number  of  fixed
             determiner s to be 4. Otherwise  set the  nodeids  of all the attributes   participat ing in
             left hand  side  of the FD in first  empty  determiner   of  this  node.
    END

Fig.5. Algorithm of adding a new FD in a relations linked list.

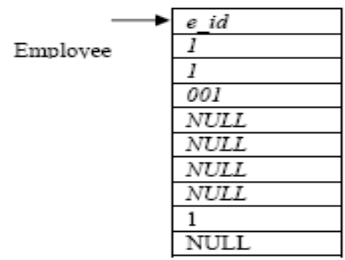

Fig. 6. Snap shot of linked list when first node is added on it.

Fig. 7 shows linked list when all the attributes are added on linked list. After adding all the attributes we need to add information about all the FD's holds on the relation Employee in the linked list representation of this relation using algorithm described in Fig. 5. Note that FD's will be added one after the other. One more thing is that we need to convert FD into a format such that right hand side will contain only dependant, this will be automatically done in finding minimal cover. Thus FD (1) will be broken into three FD's as follows

$$e\_id \rightarrow e\_s\_name$$
$$e\_id \rightarrow j\_class$$
$$e\_id \rightarrow CHPH$$

Thus we will have total 4 FD's to be added. When these four FD will be added one after the other linked list will look like as shown in Fig. 8. Not that only the *determiner of this node* fields will be updated and the *nodeid's* of their corresponding determiner are set in these fields according to algorithm shown in Fig.5.

## 6. DESIGN CONSTRAINTS.

Every system needs to be designed by taking into account set of constraints. Our system has following constraints
   1. It restricts the total number of determiners of a single dependant attribute to four. But as per knowledge of authors more frequently observed real world relations generally do not have more than four attributes as composite determiner. If implementation is done in Java then this restriction can also be removed. But if needed it can be increased.
2. It also applies restrictions on length of attribute name but by setting as much length as possible e.g. 100, any possible attribute name can be stored.
3. Order of entering the attribute can also be treated as a constraint, but it is immaterial to the user.

In overall we want to say that the constraints can easily handle most frequently observable real world relations and thus they are less restrictive.

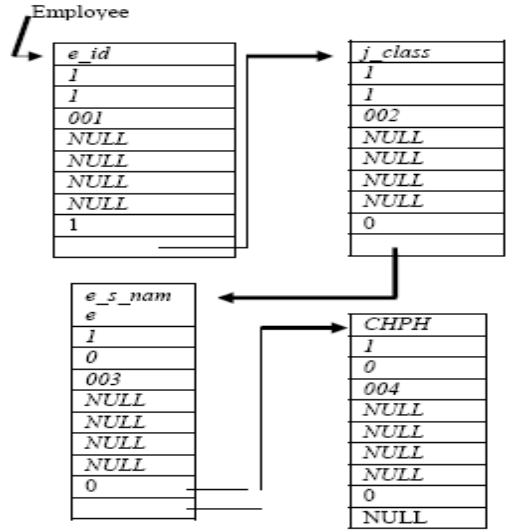

Fig. 7. Linked list when all four attributes are added

## 7. ALGORITHM OF 1NF.

Converting a relation into 1NF is done at the time of entering the relation using a GUI interface like [19]. For each composite attribute GUI asks for the set of atomic attributes corresponding to composite attribute. Thus 1NF is achieved at the time of entering the relation schema like Micro [19]. Similarly, multi-valued attributes are handled as follows. Each multi-valued attribute is replaced by "attribute name_ID", so that only one valus can be inserted at a time in that column.

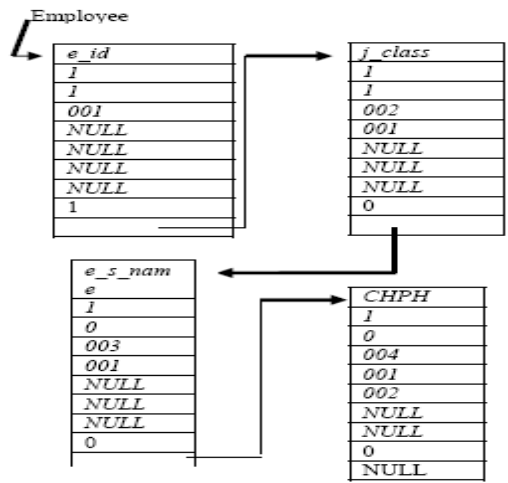

Fig. 8. Linked list after adding all FD's.

## 8. ALGORITHM OF NORMALIZATION

This algorithm takes input as a Head pointer of linked list, which stores a relation in 1NF, in computers memory in a linked list format as discussed above. Second input is a Flag3NF. Database designer will provide value of flag Flag3NF, if designer want to normalize this relation up to 3NF, one will set this flag. For normalizing this relation in 2NF, designer will reset this flag. During the process of normalization, in step 2 it creates table structures , which are nothing but array of strings and then these table structures are used to create actual tables in Oracle. This algorithm internally uses another algorithm called *AttributeInfo*, that provides PrimeAttributes[ ], AllAttributes[ ] and PrimeKeyNodeIds[ ] , which are used by remaning part of the algorithm.

*Algorithm_Normalization(Head, Flag3NF)*
{
Input: - *Head* pointer of linked list holding all the attributes and functional dependencies of
 a relation to be normalized. A flag named Flag3NF, which is set to 1 if user wants
 to normalize up to 3NF otherwise normalization will be done up to 2NF only.
Output: - If flag3NF=1 Tables created in 3NF in Oracle,
 else Tables created in 2NF in Oracle.
Let *A1*, *A2 and A3* be the string arrays used to hold the set of related attributes taking participation in full FD, partial FD and transitive dependencies (TD), respectively. A2 and A3 are divided into two components namely determiner and dependent, for storing determiner and dependent attributes participating in a given type of dependency. A2 has two components as A2-dependent[] and A2-determiner[] used for storing dependent and determiner attributes, respectively, participating in a partial FD. Similarly A3 will have two components A3-determiner[] and A3-dependent[], used for storing determiner and dependent attributes, respectively, participating in TD. Let *Listptr and Trav* are pointer variables of type structure node.

1. Calculate number of prime attributes and store attributes taking participation in different types of functional dependencies in string arrays A1, A2 and A3.
 Set *listptr=Head*;
 /*Here *Head* is a pointer variable pointing to first node of linked list.
 Call { *PrimeKeyNodeIds[ ], PrimeAttributes[ ], AllAttributs[ ]}=*A*ttributeInfo (listptr)*
 /* it returns total no of prime attributes in *KeyCount*.
 /*After execution of this algorithm we will get *node_ids* of all the prime
 /*attributes in array *primeKeyNodeId[]* and their attribute names in array
 /* *PrimeAttributes[] and list of all attributes in array AllAttributes[]*
 For (each non- key attribute) do the following
 {
 1a. Initialization.
 Set *Flag1=0 Flag2=0, Flag3=0*; *index1*=1, *index2*=1; index3=1.
 /* *index1*, index2 and *index3* are used for indexing of array *A1,* A2 and *A3,*
 /* respectively. A2-determiner[] array is used to store determiners and
 /* A2-dependant[] stores dependant attributes participating in Partial FD.
 /* Flag1, Flag2 and Flag3 are set for Full, Partial and transitive dependency,
 /* respectively.
 1b. Finding non-key attributes and their determiners for finding each type of
 dependency holds on this relation by traversing its linked list.
 Node * trav;
 *Trav = Head*;
 while(*Trav* →*ptrTonext* ≠ NULL)
 If ( *Trav* → *keyAttribute* = 0)
 Then
 Find the *determiner_ id[]* of *Trav*
 /* where *determiner_id[]* is an array of *node-id*s of all

```
                    /*the determiner attributes
             If (determiner_id[] of Trav == primeKeyNodeId[])
             Then   Set Flag1=1;  /* Full FD exists
                        /* two arrays are equal if they have exactly same elements
                  /* may be ordered in different sequence.
             End
             If (determiner_id [Trav] ⊂ primeKeyNodeId[])
                /*means partial FD exists
             Then   Set Flag2=1
             End
             /* where ⊂ is proper subset operator

             If (determiner_id [Trav] ∉ primeKeyNodeId[])
                /* where ∉ is does not belong to operator
                /*means Transitive FD exists
             Then   Set Flag3=1
             End
```

1c. Storing attributes participating in full functional dependencies in A1.

```
        If (Flag1==1)      /* means full functional dependency exists */
        Then
        /*save attribute pointed by Trav in array A1, as
```

$$A1(index1) = A1(index1) \cup (Trav \rightarrow attribute\_name) \cup \Pr imeAttributes[]$$

```
        End
        /* note that A1 will always have only one entry
```

1d. Storing attributes participating in partial functional dependencies in
    *A2-dependant* and *A2-determiner.*

```
        If (Flag2==1) /* means partial dependency exist */
        then
        /*save attributes pointed by Trav and all its determiner attributes in arrays
        /*A2-dependant and A2-determiner.
   If (determiners of this non-key attribute is already present in A2-determiner
      at k th index )
   then  A2 − dependant(k) = A2 − dependant(k) ∪ (Trav → attribute _ name)
   else  /* add a new entry in A2-dependant and A2-determiner
```

$$A2 - dependant(index2) = (Trav \rightarrow attribute\_name)$$
$$A2 - \det er\min er(index2) = (\det er\min ers \quad of \quad Trav)$$
$$index2 + +$$

```
        End
```

1e. Storing attributes participating in transitive dependencies.

```
        If (Flag3==1) /* means transitive FD exist */
        Then
        /*save attributes pointed by Trav and all its determiner attributes in arrays
        /*A3-dependant and A3-determiner
        If (determiners of this non-key attribute is already present in A3-determiner
           at k th index )
        Then  A3 − dependant(k) = A3 − dependant(k) ∪ (Trav → attribute _ name)
        else  /* add a new entry in A3-dependant and A3-determiner
```

$$A3 - dependant(index3) = (Trav \rightarrow attribute\_name)$$
$$A3 - \det er\min er(index3) = (\det er\min ers \quad of \quad Trav)$$
$$index3 + +$$

    End
}EndFor

2. Create tables in oracle
{
   This step first creates table structures T1,T21, T22,..,T2n, T31,T32,…,T3m, which are nothing but string arrays that can be directly converted to actual table definitions.
   If (*Flag3NF=0*)  /* means normalize up to 2NF only
   Then
      /*create following table structures
      2a. Create a new table structure *T1*, using union of *A1* and all entries in
          *A3-determiner* and *A3-dependent*, with primary key as all the elements in
          *PrimeAttributes[]*.
          $$T1 = A1 \cup A3-dependant \cup A3-determiner$$
      2.b. Create a separate table structures *T2i* from union of each entry of *A2-determiner* and *A2-dependent*, with primary key as element of *A2-determiner* for that entry. Let *A2* has n entries then create table structures
          as,  $T2i = A2-dependant(i) \cup A2-determiner(i)$ for $i = 1,2...,n$.

   Create tables in Oracle using table structures defined in step 2a and step 2b
   along with their primary key definitions.
   else   /* means normalize up to 3NF
      2c. Create a new table structure *T1* using all the attributes present in *A1*, with
          primary key as all the elements of *PrimeAttributes[]*.
          $$T1 = A1$$
      2d. Same as step 2b
      2e. Create separate table structures *T3i* from union of each entry of
          *A3-determiner* and *A3-dependent*, with primary key as element of
          *A3-determiner* for that entry. Let *A2* has n entries then create table
          structures  as  $T3i = A3-dependant(i) \cup A3-determiner(i)$ for
          $i = 1,2...,n$.
          Add primary key of each table structures T3i as foreign key in the table
          Structures T1, if it is not already there and update their table structures accordingly.
      Create tables in Oracle according to the table structures created and or
      updated in step 2c, 2d and 2e by defining primary and foreign keys.
   End
}

}

*AttributeInfo( Head)*
{
   Trav is variable pointer to node structure.
   *Trav=Head;*
   *AllAttributes[]=0,PrimeAttributes[]=0,PrimeKeyNodeIds[]=0;*
   While *(Trav → ptrTonext ≠ NULL)*
   {
      If *(Trav→KeyAttribute=1)*
      Then
         *PrimeAttribute[]=PrimeAttribute[] ∪ Trav→AttributeName,*
         *PrimeKeyNodeIds[]=PrimeKeyNodeIds[] ∪ Trav→nodeid,*
      Else

*AllAttributes[]=AllAttributes[] ∪ Trav→AttributeName,*
End
*}*
Return *AllAttributes[], PrimeAttributes[], PrimeKeyNodeIds[];*
*}*

## 9. EXAMPLE TRACE OF NORMALIZATION ALGORITHM

Let the relation R= {a, b, c, d, e, f, g} having the following functional dependencies hold on it. FD's ={ a,b→ c; a, b→ d; b→ e; d→ f; d→g }. Let sequence of storing attribute be a, b, c, d, e, f and g with node_id as 001, 002, 003, 004, 005, 006 and 007.therefore values of *primeKeyNodeId[],PrimeAttributes[]* will be *primeKeyNodeId=[001,002],PrimeAttributes=[a, b]*

For the first non- key attribute "c", *determiner_id[]*=[001, 002],on comparing it with P*rimeKeyNodeId=[001,002],*we will get *Flag1*=1 (Full FD) hence attribute "c" and primary key will be saved in A1as A1[1] =[a, b, c]. For the next non-key attribute "d", *determiner_id[]*=[001,002],on comparing it with *primeKeyNodeId=[001.002],*we will get *Flag1*=1 (Full FD) hence array A1 will be updated to A1[1] =[a, b, c, d]. For the third non-key attribute "e", *determiner_id[]*=[002],on comparing it with *primeKeyNodeId=[001,002],*we will get *Flag2=1* (PFD) hence A2 will be updated as *A2-determiner[1]=b, A2-dependants[1]=e.* For the fourth non-key attribute "f", *determiner_id[]*=[004],on comparing it with *primeKeyNodeId=[001.002],*we will get *Flag3=1(TD)* hence A3 will be updated as
*A3-determiner[1]=d, A3-dependants[1]=f.* For the next non-key attribute "g", *determiner_id[]*=[004],on comparing it with *primeKeyNodeId=[001.002],*we will get *Flag3=1* hence A3 will be updated as *A3-determiner[1]=d, A3-dependents[1]=[f, g], since both attributes "f" and " g" have same determiner "d"* .

After completion of step 1, contents of A1, A2 and A3 are as shown in Fig. 11.

| A1 | | A2-determiner | | A2-dependent | |
|---|---|---|---|---|---|
| | | 1 | b | 1 | e |
| 1 | a, b, c, d | A3-determiner | | A3-dependent | |
| | | 1 | d | 1 | f, g |

Fig. 11. Contents of A1, A2 and A3.

Table structures defined in step 2 if Flag3NF=0 are as follows
T1=[a, b, c, d, f, g]
T2=[b, e]
Table structures defined in step 2 if Flag3NF=1 are as follows
T1=[a, b, c, d]
T2=[b, e]
T31=[d, f, g]
Since primary key of T31, "d" is already part of T1 there is no need to add it in T1.

## 10. STANDARD RELATIONS USED FOR EXPERIMENTATION.

We want to test the performance of our tool *RDBNorma* with the existing tool Micro [19]. For this purpose we have collected 10 examples of relation normalization up to 3NF from various research papers. Table 1 shows description of these relations. Table 2 shows the decomposition of relations shown in Table 1 into 2NF and 3NF. In Table 1 and 2 FD are separated by semicolon. Table 1 is spread over multiple pages. Table 2 is used for testing the output of our tool *RDBNorma*. These relations can also be helpful to the readers as a reference.

| Sr No | Relation Name | Relation Description | No of Attributes | No of FD's |
|---|---|---|---|---|
| 1 | Beer_Relation [26] | Beer_Relation { <u>beer</u>, brewery, strength, city, region, <u>warehouse</u>, quantity }<br><br>FDs = {beer → brewery; beer → strength; brewery → city; city → region; beer, warehouse → quantity} | 7 | 5 |
| 2 | GH_Relation [21] | GH {A, B, C, D, E, F, <u>G, H</u>, I, J, K, L}<br><br>FDs = { A → B, C; E → A, D ; G → A, E, J, K ; G, H → F, I K → A, L; J → K} | 12 | 13 |
| 3 | ClientRental [27] | ClientRental { <u>clientNo, propoertyNo,</u> cName, pAddress, rentStart, rentFinish, rent, ownerNo, oName}<br>FDs = {<br>clientNo, propoertyNo → rentStart, rentFinish ;<br>clientNo → cName ; wnerNo → oName;<br>propoertyNo → pAddress, rent, ownerNo, oName;<br>clientNo, rentStart → prpoertyNo, pAddress,<br>rentFinish, rent, ownerNo, oName;<br>propoertyNo, rentStart → clientNo, cName, rentFinish; } | 9 | 17 |
| 4 | AB_Relation [21] | AB {A, B, C, D, E, F, G, H}<br><br>FDs = {A, B → C, E, F, G, H; A → D;<br>F → G; B, F → H;<br>B, C, H → A, D, E, F, G;<br>B, C, F → A, D, E; } | 8 | 16 |

| Sr No | Relation Name | Relation Description | No of Attributes | No of FD's |
|---|---|---|---|---|
| 5 | Invoice Relation [28] | Invoice( Order_ID , Order_Date, Customer_ID, Customer_Name, Customer_Address, Product_ID , Product_Description, Product_Finish, Unit_Price ,Order_Quantity ) <br><br> FDs = { Order_ID → Order_Date ; <br> Order_ID → Customer_ID; <br> Order_ID → Customer_Name; <br> Order_ID → Customer_Address; <br> Customer_ID → Customer_Name; <br> Customer_ID → Customer_Address; <br> Product_ID → Product_Description; <br> Product_ID → Product_Finish; <br> Product_ID → Unit_Price ; <br> Order_ID, Product_ID → Order_Quantity;  } | 10 | 10 |
| 6 | Emp Relation [27] | Emp  {emp_id, emp_name, emp_phone, dept_name, dept_phone, dept_mgrnname, skill_id, skill_name, skill_date, skill_lvl} <br><br> FDs = { emp_id → emp_name, emp_phone; <br> emp_id → dept_name ; <br> dept_name → dept_phone , dept_mgrnname ; <br> skill_id ← skill_name ; <br> emp_id, skill_id → skill_date , skill_lvl;  } | 10 | 8 |
| 7 | Project Relation [29] | Project {project code, project title, project manager, project budget, employeeNo, employeeName, deptNo, deptName, hourlyRate } <br> FDs = { projectCode → project title; <br> projectCode → project manager; <br> projectCode → project budget; <br> employeeNo → employeeName; <br> employeeNo → deptNo; <br> employeeNo → deptName; <br> projectCode, employeeNo → hourlyRate ; <br> deptNo → deptName;  } | 9 | 8 |

| Sr No | Relation Name | Relation Description | No of Attributes | No of FD's |
|---|---|---|---|---|
| 8 | WellmeadowsHospital [30] | WellmeadowsHospital {(Patient_No, DrugNo, Start_Date, Full_Name, Ward_No, Ward_Name, Bed_No, Drug_Name, Description, Dosage, Method_Admin, Units_Day, Finish_Date}<br><br>FDs = { Patient_No → Full_Name;<br>Ward_No → Ward_Name, Bed_No;<br>Drug_No → Drug Name, Description;<br>Drug_No → Dosage, Method_Admin;<br>Patient_No, Drug_No, Start_Date → Units_Day;<br>Patient_No, Drug_No, Start_Date → Finish_date;<br>Patient_No, Drug_No, Start_Date → Ward_No; } | 13 | 10 |
| 9 | StaffProperyInspection [27] | StaffProperyInspection(PropertyNo, idate, itime, pAddress, coments, staffNo, sName, carReg )<br>FDs = {PropertyNo, idate → itime ;<br>PropertyNo, idate → coments ;<br>PropertyNo, idate → staffNo;<br>PropertyNo, idate → sName;<br>PropertyNo, idate → carReg;<br>PropertyNo → pAddress;<br>staffNo → sName; staffNo, idate → carReg<br>carReg, iDate, iTime → PropertyNo ;<br>carReg, iDate, iTime → pAddress;<br>carReg, iDate, iTime → coments;<br>carReg, iDate, iTime → staffNo, sName;<br>staffNo, iDate, iTime → PropertyNo ;<br>staffNo, iDate, iTime → pAddress, coments ; } | 8 | 16 |
| 10 | Report [31] | Report (reportNo, editor, deptNo, deptName, deptAddress, authourId, authourName, authourAddress)<br><br>FDs = { reportNo → editor, deptNo;<br>deptNo → deptName, deptAddress;<br>authourId → authourName;<br>authourId → authourAddress; } | 8 | 6 |

Table 1. Description of standard relations used for experimentation.

## 11. EXPERIMENTAL RESULTS OF RDBNORMA.

Table 2 shows the expected output of RDBNorma collected from the above research papers. We have compared output of *RDBNorma* with the expected output from Table 2 and can say that output of *RDBNorma* is valid and it works in expected manner.

| Sr No | Relation Name | 2NF | 3NF |
|---|---|---|---|
| 1 | Beer_Database | beer {beer, brewery, strength, city, region} beerwarehouse {beer, warehouse, quantity} | beer {beer, brewery, strength} brewery {brewery, city} city{(city, region} beerwarehouse {beer, warehouse,quantity} |
| 2 | GH_Relation | GH {G, H, F, I} G {G, A, B, C, D, E, J, K, L} | GH {G, H, F, I} G {G, E, J} J {J, K} K {K, A, L} E {E, A, D} A {A, B, C} |
| 3 | ClientRental | Client (clientNo, cName ) rental(clientNo, propoertyNo, rentStart, rentFinish ) PropertyOwner(propoertyNo, pAddress, rent, ownerNo, oName) | Client (clientNo, cName ) rental l(clientNo, propoertyNo, rentStart, rentFinish ) PropertyOwner(propoertyNo, pAddress, rent, ownerNo) Owner(ownerNo, oName) |
| 4 | AB_Relation | *AB ( A, B, C, E, F, G, H )* *A (A, D)* | *AB( A, B, C, E, F, G, H )* *F (F, G)* *A (A, D)* |
| 5 | Invoice Relation | OrderLine (order_id, product_id, ordered_qty) ProductID (product_id, product_desc, product_finish, unit_price) OrderID(order_id, order_date, customer_id, customer_name, customer_address) | OrderLine (order_id, product_id,ordered_qty) Product (product_id, product_desc, product_finish,unit_price) Order(order_id, order_date, customer_id) Customer(customer_id, customer_name, customer_address ) |
| 6 | Emp Relation | empID {emp_id, emp_name, emp_phone, dept_name, dept_phone, dept_mgrnname} skill_id {skill_id, skill_name} | empID{emp_id, emp_name, emp_phone,dept_name} Dept {dept_name, dept_phone, |

International Journal of Database Management Systems ( IJDMS ), Vol.3, No.1, February 2011

| | | emp_id skill_id {emp_id, skill_id, skill_date, skill_lvl} | dept_mgrnname} SkillID{skill_id, skill_name} EMP{emp_id, skill_id, skill_date, skill_lvl} |

| Sr No | Relation Name | 2NF | 3NF |
|---|---|---|---|
| 7 | Project Relation | projectCode( ProjectCode, project title, project manager, project budget) employeeNo (employeeNo,employeeName, deptNo, deptName} projectCodeemployeeNo( projectCode, employeeNo, hourlyRate) | projectCode ( ProjectCode, project title, project manager, project budget) employeeNo (employeeNo , employeeName, deptNo) projectCodeemployeeNo( projectCode, employeeNo, hourlyRate) deptNo ( deptNo, deptName) |
| 8 | WellmeadowsHospital | Hospital ( Patient_No, Drug_No,Start_Date, Ward_No, Ward_Name, Bed_No, Units_Day, Finish_Date) Drug( Drug_No, Name, Description, Dosage, Method_Admin) Patient ( Patient_No, Full_Name ) | Hospital ( Patient _No, Drug_No, Start_Date, Ward_No, Bed_No, Units_Day, Finish_Date) Drug ( Drug_No, Name, Description, Dosage, Method_Admin ) Patient ( Patient_No, Full_Name) Ward ( Ward_No, Ward_Name ) |
| 9 | StaffProperyInspection | PropertyNo(PropertyNo, pAddress) PropertyNoidate(PropertyNo,idate,itime, coments,staffNo,sName,carReg ) | PropertyNo(PropertyNo, pAddress) PropertyNoidate(PropertyNo,idate,itime, coments,staffNo,carReg ) staffNo(staffNo, sName) |
| 10 | Report | ReportNo(report_no,editor, dept_no, dept_name, dept_addr) Authorid (author_id, author_name, author_addr) | ReportNo(report_no,editor, dept_no) DeptNo(dept_no, dept_name, dept_addr) Authorid(author_id, author_name, author_addr) |

Table 2. Shows 2NF and 3NF of standard relations taken in Table 1.

## 12. PERFORMANCE COMPARISON OF RDBNORMA WITH MICRO.

For comparison purpose both RDBNORMA and Micro [Du and Wery, 1999], are implemented using Java. We have compared performance of RDBNORMA with the Micro in terms of time required to convert a relation into 2NF and 3NF, in milli seconds. Table 3 shows this performance of Micro. Performance of proposed tool *RDBNorma*, is shown in Table 4.

The time, on average, required to convert a given relation in 2NF by Micro is around 3.9 times the time required by *RDBNorma*. The time, on average, required to convert a given relation into 3NF by Micro is 2.89 times the time required by *RDBNorma*. Thus *RDBNorma* is more faster than Micro in both 2NF and 3NF, conversions

Plot of number of attributes and time required to bring relation in 2NF and 3NF using *RDBNorma* are shown in Fig.12 and Fig.13 We have also observed that the time required to convert a given relation into 2NF and 3NF is depend not only on the number attributes of a relation but also on number of functional dependencies holds on that relation.

|  |  | 2NF Time | | 2NF Total Time | 3NF Time | | 3NF Total Time |
|---|---|---|---|---|---|---|---|
| Sr no. | Relation name | Normalization | Table Implementation |  | Normalization | Table Implementation |  |
| 1 | Beer | 567 | 698 | 1265 | 419 | 425 | 844 |
| 2 | GH | 311 | 314 | 625 | 196 | 320 | 516 |
| 3 | Client | 253 | 579 | 626 | 233 | 298 | 531 |
| 4 | AB | 339 | 380 | 719 | 215 | 456 | 671 |
| 5 | Invoice | 203 | 421 | 624 | 223 | 355 | 578 |
| 6 | EMP | 241 | 384 | 625 | 186 | 392 | 578 |
| 7 | Project | 217 | 377 | 594 | 293 | 301 | 594 |
| 8 | WMHospital | 324 | 254 | 578 | 179 | 259 | 438 |
| 9 | SPInspction | 257 | 398 | 655 | 388 | 315 | 703 |
| 10 | Report | 466 | 581 | 1047 | 249 | 362 | 611 |
| Average |  | 317.8 | 438.6 | 735.8 | 258.1 | 348.3 | 606.4 |

Table 3. Timing analysis of Micro.

|  |  | 2NF Time | | 2NF Total Time | 3NF Time | | 3NF Total Time |
|---|---|---|---|---|---|---|---|
| Sr no. | Relation name | Normalization | Table Implementation |  | Normalization | Table Implementation |  |
| 1 | Beer | 147 | 571 | 718 | 198 | 377 |  |
| 2 | GH | 175 | 309 | 481 | 190 | 295 | 485 |
| 3 | Client | 160 | 478 | 638 | 132 | 244 | 276 |
| 4 | AB | 207 | 232 | 439 | 117 | 322 | 439 |
| 5 | Invoice | 132 | 367 | 499 | 176 | 226 | 402 |
| 6 | EMP | 141 | 289 | 430 | 142 | 288 | 430 |
| 7 | Project | 129 | 305 | 434 | 193 | 245 | 438 |
| 8 | WMHospital | 205 | 243 | 448 | 116 | 230 | 346 |
| 9 | SPInspction | 165 | 377 | 542 | 132 | 309 | 441 |
| 10 | Report | 139 | 503 | 642 | 106 | 356 | 462 |
| Average |  | 160 | 367.4 | 527.1 | 150.2 | 289.2 | 429.4 |

Table 4. Timing analysis of *RDBNorma*.
.

From table 3 and 4, by comparing average time required for converting a relation in 2NF, one can conclude that, Micro needs around double time for conversion to 2NF than *RDBNorma*. Again, average time required to normalize and create a table in 2NF by Micro, is around 1.4 times than the *RDBNorma*. Time required to convert a relation from 2NF to 3NF by Micro is 1.72 times the time required by RDBNORMA. Total time needed for converting a relation from 2NF to 3NF and creating a table from them, by Micro is also 1.4 times the time required by *RDBNorma*. Thus, we can conclude that *RDBNorma* is faster than Micro.

We have also compared *RDBNorma* and Micro in terms of memory space required to store a relation in terms of number of bytes. For this purpose the above mentioned standard relations are used and the memory requirement is shown in Table 5. From Table 5 we can conclude that the memory requirement of Micro is around 2.17 times the memory requirement of *RDBNorm*

| Sr No | Relation Name | No of Attributes | No of FD's | Memory required by *Micro* in bytes | Memory required by *RDBNorma* in bytes |
|---|---|---|---|---|---|
| 1 | Beer_Relation | 7 | 5 | 14544 | 3184 |
| 2 | GH_Relation | 12 | 13 | 14752 | 7160 |
| 3 | ClientRental | 9 | 17 | 14832 | 10480 |
| 4 | AB_Relation | 8 | 16 | 14624 | 6200 |
| 5 | Invoice Relation | 10 | 10 | 14792 | 6496 |
| 6 | Emp Relation | 10 | 8 | 14776 | 9600 |
| 7 | Project Relation | 9 | 8 | 14736 | 2256 |
| 8 | WellmeadowsHospital | 13 | 10 | 14824 | 5800 |
| 9 | StaffProperyInspection | 8 | 16 | 14744 | 9520 |
| 10 | Report | 8 | 6 | 14576 | 7000 |
| | Avg | | | 14720 | 6769.6 |

Table 5. Memory requirement of Micro and RDBNORMA.

## 13. CONCLUSION

It is concluded that a relation can be represented with only one singly linked list along with its set of FD's. Thus we can save considerable space as compared with representing a relation using two linked list one for attributes and other for FD's. Since understanding linked list is easy, the representation will be easy to understand. The definitions of 2NF and 3 NF algorithms on such a representation will be efficient since linked list structure can be manipulated/accessed efficiently.

From the performance comparison of *RDBNorma* and Micro we can conclude that the time on average required to convert a given relation in 2NF by Micro is around 3.9 times the time required by *RDBNorma* and the time required to convert a given relation into 3NF by Micro is 2.89 times the time required by *RDBNorma*. Thus *RDBNorma* is at least 2.89 times faster than the Micro. When they are compared in terms of momory space, we can conclude that the memory requirement of Micro on average is around 2.17 times the memory requirement of *RDBNorma*.

Thus *RDBNorma* is better than Micro in terms of both the speed and memory space requirement.

*ACKNOWLEDGEMENT*

We are thankful to management Vishwakarma Institutes, Pune for their encouragement and whole hearted cooperation during this work. We are equally thankful to the director, Board of university and college development (BCUD), university of Pune for their motivation and guidance. First and Third authors are thankful to Prof. G.V Garje, the chairman, Board of Study – Computer Engineering, University of Pune.

## *REFERENCES*
[1] Codd E. F. (1970), "A relational model of data for large shared data banks", Communications of the ACM. vol. 13, No.6, pp. 377–387.
[2] Codd E. F. (1971), "*Further normalization of the data base relational model*", IBM Research Report, San Jose, California, vol. RJ909.
[3] Kent W.(1983), "A Simple Guide to Five Normal Forms in Relational Database Theory", Communications of the ACM. vol.26 No.2. pp.120-125.
[4] Date C. J. (1986), "An introduction to database system", fourth edition, Addison Wesley.
[5] Silberschatz, Korth and S. Sudarshan (2006), "Database system Concepts", McGraw Hill international edition, Fifth edition.
[6] Elmasri and Navathe (1994), "Fundamentals of Database systems", Addison Wesley, second edition.
[7] Ramakrishnan and Gehrke, (2003), "Database management systems", McGraw- Hill, international edition, third edition.
[8] Rob and Coronel (2001), "Database systems, design, implementation and management", Course technology, Thomson learning, fourth edition.
[9] Jui-Hsiang and Thomas C (2004), "Traditional and alternate database normalization techniques: their impact on IS/IT student's perception and performance", International Journal of information technology education, vol 1, No. 1, pp.53-76.
[10] Salzberg B (1986), "Third normal form made easy", SIGMOD record, Vol.15, No.4, pp. 2-18.
[11] Ling T, Tompa F. W. and Kameda T (1981), "An improved 3NF", ACM Transactions on Database Systems, Vol.6, No.2, pp.329-346.
[12] Beeri C and Bernstein P. A.(1979), "Computational problems related to the design of normal form relational schemas", ACM Transactions on Database Systems, Vol.4, No.1, pp.30-59.
[13] Fagin R (1977), "Multivalued dependencies and a new normal form for relational databases", ACM Transactions on Database Systems, Vol.2, No.3, pp.262-278.

[14] Fagin R (1979), "Normal forms and relational database operators", ACM SIGMOD International Conference on Management of Data, Boston, Mass., pp. 153-160.
[15] Fagin R (1981), "A normal form for relational databases that is based on domains and keys", ACM Transactions on Database Systems, Vol.6, No.3, pp.387- 415.
[16] Zaniolo C (1982), "A new normal form for the design of relational database schemata", ACM Transactions on Database Systems, Vol.7, No.3, pp.489- 499.
[17] Diederich J and Milton J (1988), "New methods and fast algorithms of database normalization", ACM transactions on database System, Vol.13, No.3, pp. 339-365.
[18] Hetch and Stephen C (1998), US Patent 5778375 - Database normalizing system.
[19] Du H and Wery L (1999), " Micro: A normalization tool for relational database designers", journal of network and computer application, Vol.22, pp.215-232.
[20] Yazici A, Ziya K (2007), "JMathNorm: A database normalization tool using mathematica", In proc. international. conference on computational science, pp.186-193.
[21] Bahmani A, Naghibzadeh M and Bahmani B (2008) , "Automatic database normalization and primary key generation", Niagara Falls Canada IEEE.
[22] Maier D (1988), "The Theory of relational databases", Computer science press: Rockville, MD.
[23] Antony S. R. and Batra D (2002), "CODASYS: A consulting tool for novice database designers", ACM SIGMIS, vol.33, issue 3, pp.54-68.
[241 Georgiev Nikolay (2008), "A web based environment for learning normalization of relational database schemata", masters thesis, Umea university, Sweden.
[25] Kung Hsiang-Jui and Tung Hui-Lien (2006), "A web based tool to enhance teaching/Learning database normalization", in Proceeding of international conference of southern association for information system.
[26] http://www.cs.man.ac.uk/horrocks/Teaching/cs2312/Lectures/PPT/NFexamples.ppt
[27] Thomas C and Carolyn B (2005), "Database Systems", Pearson, third edition.
[28] http://www.cs.gmu.edu/ aobaidi/spring-02/Normalization.ppt
[29] https://sta_.ti.bfh.ch/erj1/Datenbank1/slides/db05Normalizationnew2.pdf
[30] O'nell P and O'nell E (2001), "Database Principles Programming and Performance", Harcourt, second edition.
[31] Teorey T. J.(2002), Database Modeling and Design", Harcourt, third edition.

**Authors**

Y. V. Dongare (yashwant_dongre@yahoo.com), has completed ME (computer science and engineering.) in year 2009 from Vishwakarma Institute of Technology, Pune and BE (computer science and engineering.) in year 2003 from Shree Guru Govind Singhji College of engineering and Technology, Nanded, Maharashtra, India He is presently working as assistant professor in the department of computer engineering at Vishwakarma Institute of Information Technology, Pune, Maharashtra state, India. His area of interest includes database normalization, software engineering and development technologies (J2se/J2ee).

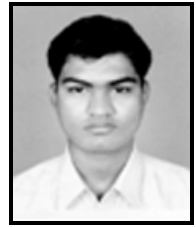

P. S. Dhabe (dhabeps@gmail.com), has completed ME in computer technology from S.G.G.S college of engineering and technology, Nanded, Maharashtra, India in year 2002. He is presently perusing his Ph.D. in systems and control engineering, IIT Mumbai. He is presently working as assistant professor in computer engineering at Vishwakarma Institute of technology, Pune from 2005. He is principal investigator of research project funded by university of pune, Maharashtra, India, titled "Database normalization tool". His areas of interest are database normalization, fuzzy neural networks and pattern recognition.

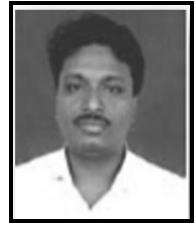

S. V. Deshmukh (supriya_vd2005@yahoo.com), has completed BE (computer engineering) in year 2008 from Vishwakarma Institute of Information Technology, Pune. She is presently working as Lecturer in the department of computer engineering at Pune Vidhyarthi Griha's College of Engineering & Technology , Pune, Maharashtra state, India. Her area of interest includes DBMS, Software Testing Quality Assurance and OOP Methodologies.

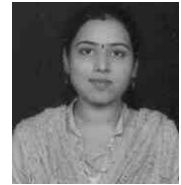